\documentclass[letterpaper, 10 pt, conference]{ieeeconf}

\IEEEoverridecommandlockouts
\usepackage{cite}
\usepackage{amsmath,amssymb,amsfonts}
\usepackage{algorithmic}
\usepackage{graphicx}
\usepackage{textcomp}
\usepackage{xcolor}
\usepackage{siunitx}

\usepackage{threeparttable}
\usepackage{booktabs}

\usepackage{acronym}
\usepackage{subcaption}
\usepackage{float}
\usepackage{multirow}
\usepackage{soul}
\usepackage{comment}
\usepackage{makecell} 

\usepackage[hyphens]{url}

\usepackage[ruled,%
            linesnumbered]{algorithm2e}

\usepackage{xcolor}

\usepackage[acronym]{glossaries}

\usepackage{hyperref}

\usepackage{amssymb}
\usepackage{pifont}
\usepackage{array}
\newcolumntype{P}[1]{>{\centering\arraybackslash}p{#1}}
\newcolumntype{M}[1]{>{\centering\arraybackslash}m{#1}}

\def\BibTeX{{\rm B\kern-.05em{\sc i\kern-.025em b}\kern-.08em
    T\kern-.1667em\lower.7ex\hbox{E}\kern-.125emX}}

\makeatletter
\def\ps@IEEEtitlepagestyle{%
  \def\@oddhead{}
  \def\@evenhead{}%
  \def\@oddfoot{%
    \vbox to0pt{\vss
      \hbox to\textwidth{%
        \parbox[t]{\textwidth}{\centering\scriptsize
          \copyright 2025 IEEE.  Personal use of this material is permitted.  Permission from IEEE must be obtained for all other uses, in any current or future media, including reprinting/republishing this material for advertising or promotional purposes, creating new collective works, for resale or redistribution to servers or lists, or reuse of any copyrighted component of this work in other works.
        }%
      }%
    }%
  }%
  \def\@evenfoot{}%
}
\makeatother
\begin{document}

\acrodef{HR}{Heart Rate}
\acrodef{BLE}{Bluetooth Low Energy}
\acrodef{DSP}{Digital Signal Processing}
\acrodef{DD}{Drowsiness Detection}
\acrodef{NN}{Neural Network}
\acrodef{EEG}{Electroencephalography}
\acrodef{FFC}{Flexible Flat Cable}
\acrodef{ULP}{Ultra Low Power}
\acrodef{IMU}{Inertial Measurement Unit}
\acrodef{FR}{frame rate}
\acrodef{PMIC}{Power Management Integrated Circuit}
\acrodef{WULPUS}{Wearable Ultra Low-Power Ultrasound}
\acrodef{US}{Ultrasound}
\acrodef{ULP}{ultra low-power}
\acrodef{AFE}{analog front-end}
\acrodef{RF}{radio frequency}
\acrodef{ML}{machine learning}
\acrodef{FPGA}{Field-Programmable Gate Array}
\acrodef{LVDS}{low-voltage differential signalling}
\acrodef{SPI}{serial peripheral interface}
\acrodef{MCU}{microcontroller unit}
\acrodef{ADC}{analog to digital Converter}
\acrodef{HDL}{hardware description language}
\acrodef{DMA}{direct memory access}
\acrodef{HMI}{Human-Machine Interface}
\acrodef{HD}{hardware design}
\acrodef{AD}{algorithm development}
\acrodef{FPS}{frames per second}
\acrodef{DAS}{delay-and-sum}
\acrodef{TGC}{time-gain compensation}
\acrodef{FP}{floating point}
\acrodef{BW}{bandwidth}

\acrodef{FFT}{Fast Fourier Transform}
\acrodef{SSVEP}{Steady State Visually Evoked Potential}
\acrodef{BCI}{Brain-Computer Interface}
\acrodef{PULP}{Parallel Ultra Low Power}
\acrodef{SoC}{System on Chip}
\acrodef{BLE}{Bluetooth Low Energy}
\acrodef{EEG}{electroencephalography}
\acrodef{EMG}{electromyography}
\acrodef{ECG}{electrocardiogram}
\acrodef{PPG}{Photoplethysmogram}
\acrodef{LOSO}{leave-one-subject-out}
\acrodef{CV}{cross validation}
\acrodef{SoA}{state-of-the-art}
\acrodef{HRV}{heart rate variability}


\newcommand{\review}[1]{\textcolor{black}{#1}}

\title{\LARGE \bf
Real-Time, Single-Ear, Wearable ECG Reconstruction, R-Peak Detection, and HR/HRV Monitoring
}

\author{Carlos Santos$^{1}$, Sebastian Frey$^{1}$, Andrea Cossettini$^{1}$, Luca Benini$^{1,2}$, Victor Kartsch$^{1}$}

\maketitle
\thispagestyle{IEEEtitlepagestyle}  
\pagestyle{empty}                   

\begingroup
  \renewcommand\thefootnote{}
  \footnotetext{%
  $^{1}$Carlos Santos, Sebastian Frey, Andrea Cossettini, Luca Benini, and Victor Kartsch are with the Integrated Systems Laboratory, ETH Z{\"u}rich, Z{\"u}rich, Switzerland {\tt\small \{csantos, sefrey, cosandre, lbenini, victor.kartsch\} @ethz.ch}\\ %
 $^{2}$Luca Benini is also with the DEI, University of Bologna, Bologna, Italy {\tt\small luca.benini@unibo.it}\\
    The authors acknowledge support from the Swiss National Science Foundation (Project PEDESITE) under grant agreement 193813.%
  }
\endgroup


\begin{abstract}
Biosignal monitoring, in particular heart activity through heart rate (HR) and heart rate variability (HRV) tracking, is vital in enabling continuous, non-invasive tracking of physiological and cognitive states. Recent studies have explored compact, head-worn devices—such as earbuds—for HR and HRV monitoring to improve usability and reduce stigma. However, this approach is challenged by the current reliance on wet electrodes, which limits usability outside clinical settings, the weakness of ear-derived signals, making HR/HRV extraction more complex, and the incompatibility of current algorithms with embedded deployment. 
This work introduces a single-ear wearable system for real-time ECG (Electrocardiogram) parameter estimation, which directly runs on BioGAP, an energy-efficient device for biosignal acquisition and processing. By combining SoA in-ear electrode technology, an optimized DeepMF algorithm, and BioGAP, our proposed subject-independent approach allows for robust extraction of HR/HRV parameters directly on the device with just 36.7~uJ/inference at comparable performance with respect to the current state-of-the-art architecture, achieving 0.49 bpm and 25.82 ms for HR/HRV mean errors, respectively and an estimated battery life of 36h with a total system power consumption of 7.6~mW. 

\textit{Clinical relevance--} The ability to reconstruct ECG signals and extract HR and HRV paves the way for continuous, unobtrusive cardiovascular monitoring with head-worn devices. In particular, the integration of cardiovascular measurements in everyday-use devices (such as earbuds) has potential in continuous at-home monitoring to enable early detection of cardiovascular irregularities.
\end{abstract}

\vspace{-0.2cm}
\section{Introduction}
\vspace{-0.1cm}

Biosignal monitoring has become a key focus of research and development, spurred by a growing interest in tracking health parameters to improve quality of life \cite{dunn2018wearables}. This field holds transformative potential by enabling disease prevention and management, enhancing athletic and occupational performance, and facilitating the monitoring of metrics such as sleep, stress, and energy levels through innovative, user-friendly wearable devices and sensors\cite{matthews2023advances, sarhaddi2022comprehensive, datwyler_softpulse}. With the advent of the Internet of Things (IoT) and efficient on-device processing platforms, wearables can now also extract and analyze complex signal patterns locally \cite{gapses}.

Heart activity monitoring is a cornerstone for various physiological states, where \ac{HR} and \ac{HRV} provide valuable insights into cardiovascular health and performance. HR measures the frequency of heartbeats per minute, while HRV reflects the variation in time intervals between successive heartbeats. In clinical settings, heart activity is commonly recorded using the Lead I setup, which measures the electrical potential difference between electrodes placed on the right and left arm. Recent advancements in wearable technology have demonstrated the feasibility of capturing electrocardiogram (ECG) signals from the ear, utilizing the so-called in-ear ECG\cite{vonRosenbergWilhelm2017Hfor, 8857547} that could be captured via 'hearable' devices such as earbuds. However, this approach presents unique challenges due to a reduced signal-to-noise ratio (SNR) caused by the smaller inter-electrode distance and susceptibility to interference from other locally produced biosignals, such as Electroencephalography (EEG), Electrooculography (EOG), and Electromyography (EMG).

Recent advancements in signal processing and machine learning \cite{DaviesHarryJ.2024TDFR} have demonstrated the feasibility of reliably extracting ECG parameters from in-ear electrode pairs (right to left ears) in real time. However, to drive widespread adoption of this technology, remaining challenges need to be addressed, namely, transitioning to dry electrodes, reducing the system complexity, and embedding algorithms directly on the wearable itself efficiently (optimization must preserve resources for primary applications like audio playback while minimizing any impact on battery life) \cite{guermandi2022wireless}.

Our work addresses the challenges above by presenting a fully-dry, single-ear wearable system for real-time embedded HR and HRV parameter estimation. This is achieved by integrating advanced in-ear dry electrode technology \cite{datwyler_softpulse}, an energy-efficient DeepMF algorithm implementation\cite{DaviesHarryJ.2024TDFR}, and BioGAP, a \ac{SoA} embedded platform for biosignal acquisition and processing. The primary contributions of this work are as follows:

\begin{itemize} 
    \item First-time demonstration of BioGAP, a \ac{SoA} platform for ultra-low-power biosignal processing, for in-ear wearable ECG measurements.
    \item Design of an optimized DeepMF-based R-peak detection (DeepMF-mini) with improved memory footprint while retaining SoA accuracy. Code and data are available under \href{https://github.com/pulp-bio/DeepMF-mini}{https://github.com/pulp-bio/DeepMF-mini}.
    \item Demonstration of user-generic ear-ECG reconstruction and HR/HRV measurements.
    \item Comparative analysis of ear-to-ear and single-ear measurements, demonstrating that the single-ear setup, which reduces setup complexity, achieves SoA performance for HV and HVR.
    \item Deployment of the complete system on BioGAP, with an inference time of 0.95 ms and a battery life of 18~h.
\end{itemize}
\vspace{-0.2cm}
\section{Related Work}
\label{sec:methods}
\vspace{-0.1cm}

In the context of 'hearable' technology for ECG monitoring, von Rosenberg \textit{et al.} \cite{vonRosenbergWilhelm2017Hfor} investigated the use of head and in-ear biopotential measurements taken symmetrically across the face's sagittal plane, demonstrating a high correlation between these signals and the standard Lead I ECG. Recently, Yarici \textit{el at.} \cite{YariciMetin2024Hfor} drew similar conclusions from electrical conduction studies taken on a single side of the face and ear locations, pointing to the feasibility of single-ear ECG monitoring. Despite being the only work (to the best of our knowledge) to also explore single-ear measurements, \cite{YariciMetin2024Hfor} only provides a feasibility study without real-time R-peak detection. Here we overcome these limitations by demonstrating real-time detection of R-peaks and a quantitative assessment of HR/HRV.
Recent research has aimed at automatizing R-Peak detection for real-time physiological state characterization, mostly with cross-ear experimental setups. In \cite{ChanwimalueangTheerasak2015ERdi}, authors presented the Matched Filter Hilbert Transform (MF-HT) algorithm denoting a superior performance to existing methods for face-lead signals with low SNR. Additionally, measuring HRV directly from across-the-head ear biopotentials was explored in \cite{TianHaozhe2023HHRV}, evaluating the benefits of ear-PPG signals to extract breath and physiological states. However, these results are based on a non-wearable data acquisition platform. Instead, here we demonstrate an increased R-peak detection accuracy (95\%, compared to the 68\% of \cite{TianHaozhe2023HHRV}) with a fully wearable platform.
With the advent of Deep Learning, authors in \cite{DaviesHarryJ.2024TDFR} proposed the Deep Matched Filter (DeepMF) architecture. First trained as an encoder-decoder to extract a latent space representation of the ear biopotential (left ear to right ear, correlated with the Lead I ECG)
, it then replaces the decoder for a classifier that is able to detect the peak locations with precision/recall as high as 95\%/92\%. 
In this work, we build on the results of \cite{DaviesHarryJ.2024TDFR}, introducing a much lighter model (more than one order of magnitude smaller, compatible with the constraints of an ultra-low-power processor for in-ear devices) without compromising detection accuracy. Also, we extend the analyses to single-ear measurements (which entail much lower SNR).
Finally, Occhipinti \textit{et al.} \cite{OcchipintiEdoardo2024IESE} presented a 1D denoising convolutional autoencoder (DCAE) to map the noisy ear-ECG to the corresponding Lead I ECG signal, achieving SoA results in HR estimations (error of 4.52 bpm). Our work improves upon these results by reporting an improved HR error (0.49 and 5.91 bpm for for cross-ear and single-ear setup respectively) based for the first time on a fully-wearable setup.

In summary, while many recent algorithmic developments are designed for real-time implementation, their deployment on embedded devices remains limited due to computational complexity and embedded platform constraints, such as memory and processing capabilities. Among the available embedded platforms for bioprocessing,
BioGAP, a compact, ultra-low-power multi-modal acquisition and processing platform, offers SoA performance compatible with AI-driven applications\cite{frey2023BioGAP}. Combined with advancements in dry electrode technologies\cite{datwyler_softpulse}, these innovations pave the way for fully embedded devices capable of unobtrusive and efficient biosignal extraction and are the base for this research. 

Table \ref{table:ear_ecg_comparison} compares our work to SoA, and a more detailed discussion is presented later in Sect. \ref{soa_comparison}.
Compared with the previous works, our implementation demonstrates, for the first time, the feasibility of real-time R-peak detection and HR/HVR estimation based on a single-ear setup, which significantly helps reduce hardware system complexity and intrusiveness. Furthermore, thanks to our optimizations of the DeepMF algorithm that leads to a 75$\times$ memory reduction, the proposed system can efficiently be deployed in a wearable platform, allowing for a battery life of 36h.

\vspace{-0.1cm}
\section{Materials and Methods}
\label{sec:materials_and_methods}
\vspace{-0.1cm}

\label{subsec:overview}

\begin{figure*}[tbh]
\centering
\includegraphics[width=0.98\linewidth, trim=0cm 0cm 0cm 0cm, clip]{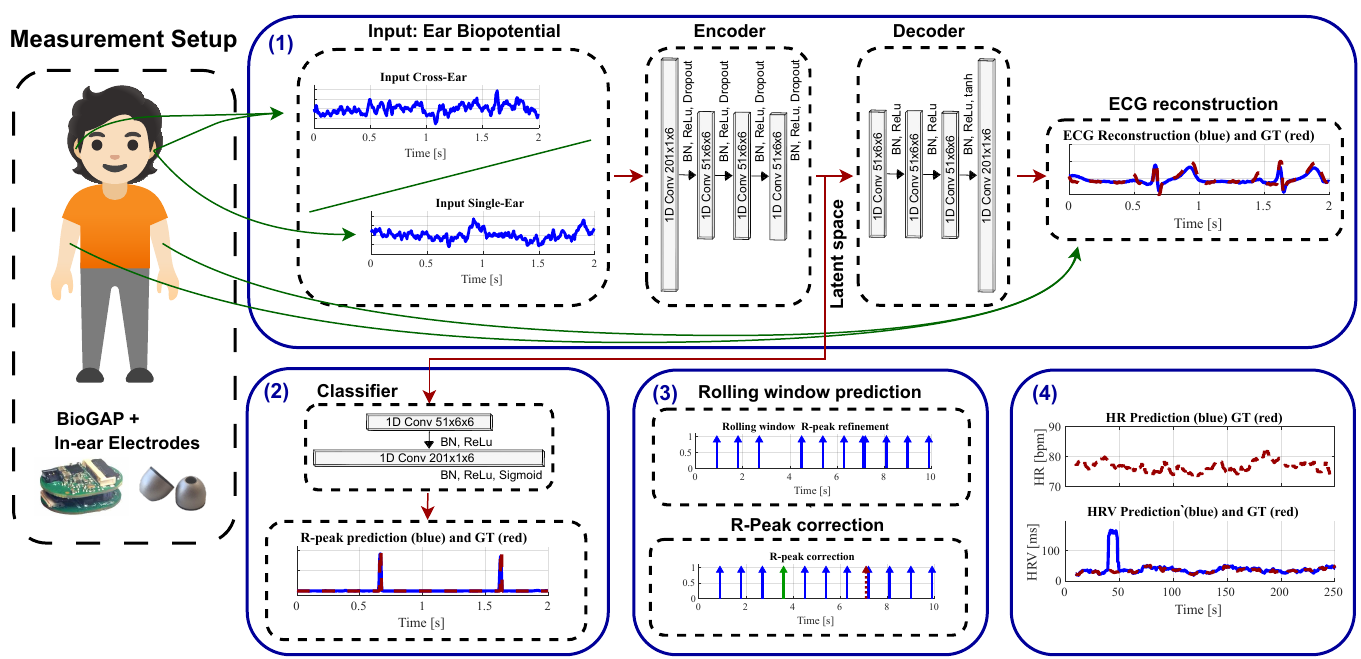}
\vspace{-0.1cm}
\caption{Overview of the measurement setup (left) and algorithmic flow: (1) autoencoder training, (2) R-peak detection classifier, (3) rolling-based postprocessing for improved R-peak detection, (4) Real-time HR/HRV calculation.}
\label{fig:prototype}
\vspace{-0.5cm}
\end{figure*}

We propose \textit{DeepMF-mini}, a lighter variant of the DeepMF \cite{DaviesHarryJ.2024TDFR} architecture with significantly fewer model parameters. We demonstrate the application of DeepMF-mini as a user-generic model for real-time R-peak detection and HR/HRV estimation with both cross-ear and single-ear setups.
Fig.~\ref{fig:prototype} presents an overview of our experimental setup and signal processing methodology, which is discussed more in detail in the following sub-sections.

\vspace{-0.1cm}
\subsection{Data collection setup}
\label{subsec:data_collection}
\vspace{-0.1cm}
We collect a dataset of simultaneous arm-ECG and in-ear biopotential measurements from four healthy volunteers (4 male, mean age: $31.0 \pm 5.1$ years). The Institution’s Ethical Review Board approved all experimental procedures, and all participants provided informed consent prior to their participation. All the experimental procedures followed the principles outlined in the Helsinki Declaration of 1975, revised in 2000.

In-ear biopotential measurements are based on two ear-EEG electrodes (SoftPulse, Dätwyler Schweiz AG), one per ear. The electrodes are made from black electrically conductive rubber coated with a silver-silver chloride layer and are ergonomically shaped to resemble consumer electronics in-ear headphones. A local buffering stage at the electrodes reduces noise and mitigates high skin contact impedance, ensuring reliable signal acquisition. In-ear electrodes are always worn dry (no conductive gel used).
Reference and bias (wet) electrodes are positioned on the left and right mastoids. 

The electrodes are connected to BioGAP \cite{frey2023BioGAP}, a compact device (16x21x14~mm$^3$, 6~g) designed for biosignal acquisition (ADS1298, Texas Instruments), ultra-low-power embedded processing (multi-core GAP9 SoC, from GreenWaves Technologies) and Bluetooth Low Energy connectivity (nRF52811, Nordic Semiconductor).

Arm-ECG measurements (ground truth) are collected by positioning two standard wet electrodes on the left and right upper arms. BioGAP collects both arm and ear signals concurrently (hence, signals are always synchronized).

This setup enables two types of measurements:
\textit{single-ear} (which captures the depolarization within the left ear region by measuring the left in-ear electrode to the reference on the left mastoid) and \textit{cross-ear} (which captures the depolarization across the head by measuring the right in-ear electrode to the reference on the left mastoid).

The collected dataset consists of four sessions, each lasting five minutes, with complete removal and re-application of the electrodes between sessions (to mimic real-life use). Measurements are performed in rest conditions: during each session, participants were instructed to remain still (no specific measures were taken to eliminate residual motion artifacts).

\vspace{-0.2cm}
\subsection{Data Preparation and Pre-processing}
\vspace{-0.1cm}
All signals (ground truth ECG and ear ExG) are always first preprocessed using a 6th-order IIR Butterworth notch filter centered at \SI{50}{Hz} to eliminate electrical interference, followed by a 2nd-order IIR Butterworth bandpass filter with cutoff frequencies of 0.5 Hz to 30 Hz to isolate the relevant ECG frequency components. 
With all signals filtered, the ground truth is computed by subtracting the right-arm signal from the left-arm signal, normalizing (z-score, zero mean, unit variance), and downsampling the signal to \SI{250}{Hz}. A discrete derivative is then calculated to emphasize the peaks, and the R-Peak locations are extracted by identifying local maxima under the constraints of a minimum peak height of 0.7 and minimum inter-peak distance of 80 samples. As labels for training the classifier, we create a binary vector to indicate the ECG R-peak locations. The values are set to 1 at the indices corresponding to the R-peak locations and at the two neighboring indices. All other values are set to 0. Both cross-ear and single-ear signals do not require further preprocessing after filtering.

In the following, unless otherwise stated, the biopotentials (ear-ExG and arm-ECG) are processed in windows with a duration of 2-seconds and a 90\% overlap.

\subsection{Training: autoencoder}
\label{subsec:r_peak_prediction}

As a first approximation to the problem, a convolutional encoder-decoder is trained to map ear biopotentials, derived from a two-second window of filtered signals recorded either from a single ear or across ears, to their simultaneous arm-ECG traces. The reconstruction target for this process is the ground truth arm ECG waveform. Through a series of 1D convolution operations, the encoder block transforms the input ear biopotentials into a latent space representation that effectively captures the shared information between the ear-ECG and the reference arm-ECG. The decoder then reconstructs this information into a waveform that resembles the arm-ECG. The encoder-decoder architecture is inspired by the DeepMF architecture, but a tanh activation is added at the end of the Decoder, which helps mitigate subject-to-subject variability in the ground truth ECG and has been empirically shown to reduce reconstruction error, thereby improving the overall performance of the encoder-decoder model.
We train the model with a user-generic approach using \ac{LOSO} \ac{CV}

\vspace{-0.1cm}    
\subsection{Training: R-peak detection classifier}

Once the encoder is successfully trained, its weights are frozen, and the decoder is replaced by a classifier designed to detect peak locations from the latent space representation generated by the encoder. The classifier is trained in LOSO CV using as ground truth the binary vector of ones and zeros described above. 
At each step, the model outputs a 1D probability vector over the 2-second window, where values close to 1 indicate high confidence in the presence of an R-peak.
Unlike the DeepMF architecture, which uses fully connected layers in its classifier, we use transpose convolutions, significantly reducing the number of model parameters from 759k to less than 10k while maintaining model performance.

\vspace{-0.1cm}
\subsection{Inference: real-time rolling-based R-peak postprocessing}

\begin{figure}[tbh]
\centering
\includegraphics[width=1\columnwidth, trim=0cm 0cm 0cm 0cm, clip]{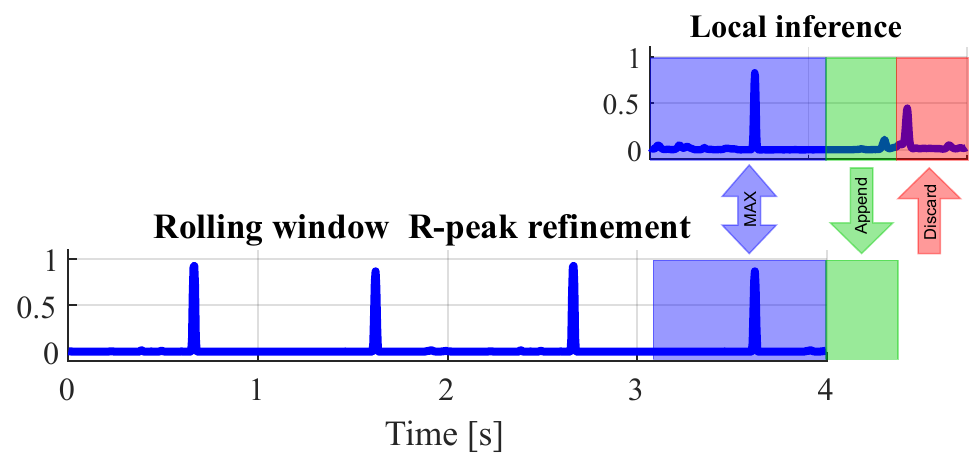}

\vspace{-0.2cm}
\caption{Rolling window R-peak refinement process.}
\label{fig:output_refinement}
\vspace{-0.3cm}
\end{figure}

Once the DeepMF-mini model is trained, we postprocess its predictions with a rolling window approach with a 2-second window size and a \SI{0.4}{s} shift per step. This overlap ensures that each segment is evaluated multiple times, increasing the likelihood of correctly identifying R-peaks.  

First, since the final \SI{0.4}{s} of each window is less reliable (as cardiac cycles corresponding to R-peaks near the boundary may not be cut off at the end of the prediction window, leading to potential misclassification), we discard the last \SI{0.4}{s} of each window (see Fig.~\ref{fig:output_refinement}, red part)

Then, we compute the element-wise maximum between the previous window inference (containing the probability of having R-peaks, as described above) and the newly obtained inference within their \SI{1.2}{s} overlap (see Fig.~\ref{fig:output_refinement}, blue part - where the 1.2 second overlap derives from having 2 second windows, 0.4 s shift and 0.4 s of end-of-window removal). This approach ensures strong detections across multiple evaluations. Then, the next \SI{0.4}{s} segment is appended, and we repeat the rolling processing (as depicted in  Fig.~\ref{fig:output_refinement}, green part).

Finally, a correction algorithm (Algorithm~\ref{algo:correct_HR}) adjusts the detection by inserting missed peaks and removing duplicates, using the running median of past RR intervals. The refined peaks are then stored in a \SI{10}{s} window for further processing.

\begin{algorithm}
\footnotesize
  \SetKwData{PredPeaks}{prediction\_peaks}
  \SetKwData{RRPredInt}{RR\_pred\_intervals}\SetKwData{RRCorrInt}{RR\_corr\_intervals}
  \SetKwData{RRMedian}{RR\_median}\SetKwData{Extra}{n\_extra}
  \SetKwData{Missed}{n\_missed}
  \SetKwData{Expects}{expectations}
  \SetKwData{Expect}{expectation}
  \SetKwData{HRVCorr}{HRV\_corr}\SetKwData{HRCorr}{HR\_corr}
  \SetKwFunction{Round}{round}\SetKwFunction{ContainZero}{contain\_0}
  \SetKwFunction{Diff}{1D\_diff}\SetKwFunction{Median}{median}
  \SetKwFunction{CorrectExtra}{correct\_extra}\SetKwFunction{Copy}{copy}
  \SetKwFunction{GetHR}{get\_HR}\SetKwFunction{GetHRV}{get\_HRV}
  \SetKwInOut{Input}{input}\SetKwInOut{Output}{output}
  \SetKw{In}{in}\SetKw{True}{True}

  \Input{\PredPeaks}
  \Output{\HRVCorr, \HRCorr}
  \BlankLine
    
  \RRPredInt $=$ \Diff{\PredPeaks}\;
  \RRCorrInt $=$ \Copy{\RRPredInt}\;
  \RRMedian $=$ \Median{\RRCorrInt}\;
  \Expects = \Round{$\frac{\RRCorrInt}{\RRMedian}$}\;
  \Extra, \Missed $= 0$
  
  \While{\ContainZero{\Expects} $==$ \True}{
    \RRCorrInt = \CorrectExtra{\RRCorrInt} \;
    \RRMedian $=$ \Median{\RRCorrInt}\;
    \Expects = \Round{$\frac{\RRCorrInt}{\RRMedian}$}\;
    \Extra $+= 1$\;
  }
  \For{\Expect in \Expects $!= 1$}{
    \Missed $+= \Expect$ \;
  }
  \RRCorrInt $=$ $\frac{\RRCorrInt}{\Expects}$\;
  \caption{R-peak  correction}\label{algo:correct_HR}
\end{algorithm}

\vspace{-0.3cm}
\subsection{Inference: \ac{HR} and \ac{HRV} calculation}

\ac{HR} and \ac{HRV} are computed based on the detected R-peaks within the most recent \SI{10}{s} window. HR is calculated by scaling the number of detected peaks in this period to a 1-minute duration, as defined in Equation~\ref{eq:HR}.

\begin{equation}
HR(\textrm{bpm}) = n_{\textrm{peaks}} \cdot \frac{60}{\frac{\textrm{last}_{\textrm{peak}} - \textrm{first}_{\textrm{peak}}}{f_s}}
\label{eq:HR}
\end{equation}

\ac{HRV} is determined as the root mean square of successive RR interval differences (RMSSD), providing a measure of beat-to-beat variability. The RR interval refers to the time between consecutive R-peaks in the ECG signal, representing the variability in heartbeats. In our analysis, HRV is expressed in milliseconds (ms) and is computed as follows (Equation~\ref{eq:HRV}):

\begin{equation} HRV (ms) = \sqrt{\tfrac{\sum_{i=1}^{n-1}(RR_{i+1}-RR_{i})^2 }{n-1} } \label{eq:HRV} \end{equation}

If the number of R-peaks detected is less than four in the computing time interval (10 s), HR and HRV are set to 0.

\subsection{Accuracy evaluation}

We evaluate precision, recall, and F1-score comparing the detection from the ear-ExG to the arm-ECG ground truth. To minimize subject bias in the \ac{LOSO} evaluation, the threshold maximizing the F1-score was selected for rolling output inference across all subjects.

\subsection{Edge Deployment}
\label{subsec:edge_deployment}

The edge deployment of our model targets the GAP9 microcontroller. We export the trained model in Open Neural Network Exchange (ONNX) format and load it using NNTool \cite{nntool}, which facilitates porting NN graphs to the GAP processor. NNTool applies model optimizations, 8-bit post-training quantization, and C code generation with optimized memory movement between internal and external memory, minimizing access overhead.

We evaluate the network’s performance on the BioGAP platform in two configurations:
\begin{itemize}
    \item \textbf{Maximum inference speed} – Running the NE16 at \SI{370}{MHz}, \SI{0.8}{V} for the fastest execution.
    \item \textbf{Highest energy efficiency} – Running at \SI{240}{MHz}, \SI{0.65}{V}, reducing power consumption quadratically.
\end{itemize}
\section{Results}

\vspace{-0.1cm}

\subsection{Signal Characterization}

\begin{figure}[tbh]
\centering
\includegraphics[width=0.85\columnwidth, trim=0cm 0cm 0cm 0.3cm, clip]{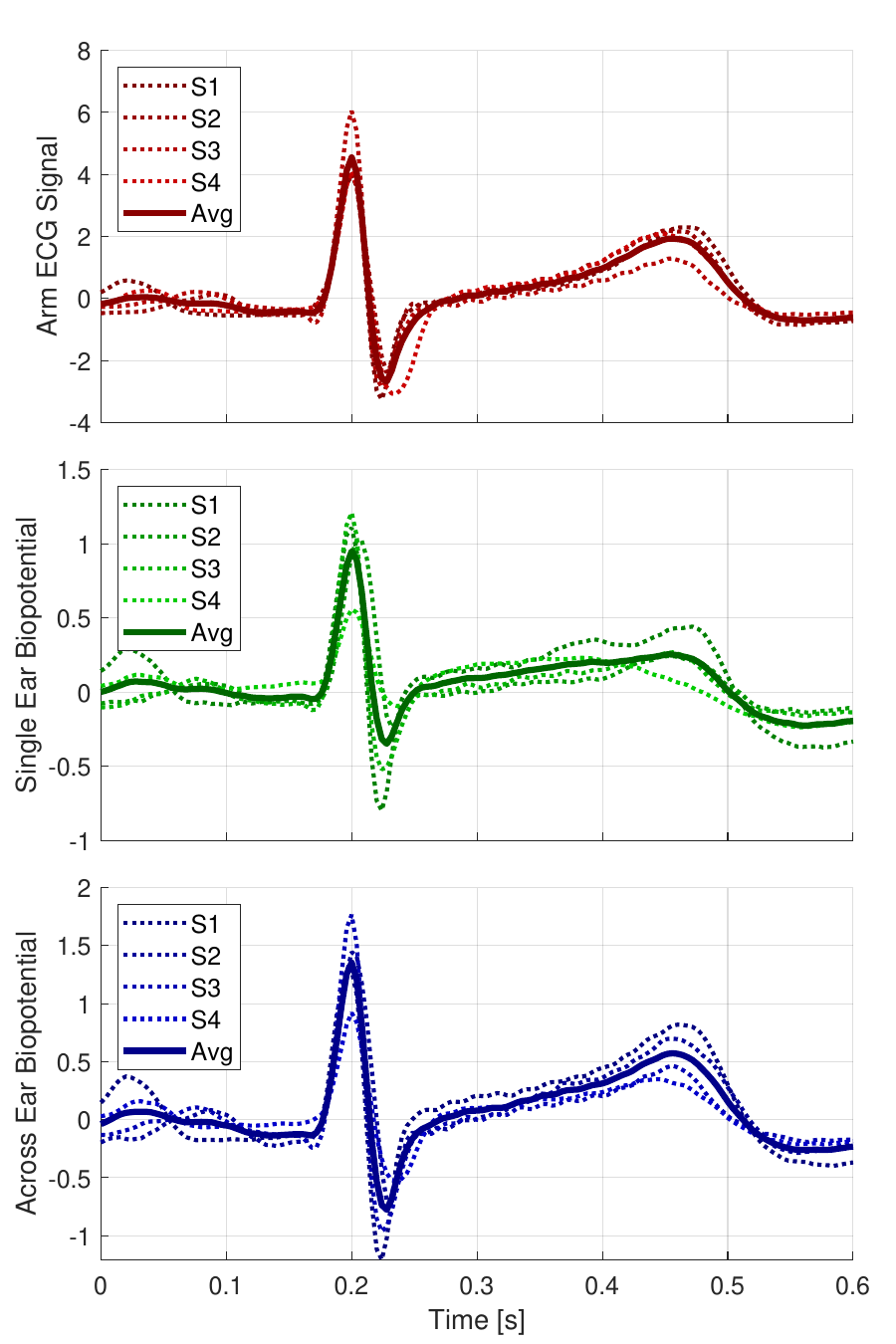}
\caption{Biopotential measurements aligned to the ground truth R-peaks and averaged over all recordings for the arm ECG (top), single-ear biopotential (middle), and cross-ear biopotential (bottom).}
\label{fig:average_ECG}
\vspace{-0.5cm}
\end{figure}

Fig.~\ref{fig:average_ECG} presents the average reconstructed ECG waveform for each subject, along with the global mean, all aligned to the detected R-peaks of the arm-ECG. The ground truth ECG (top, red) distinctly highlights the QRS complex and T wave across all subjects. The single-ear biopotential (middle, green) and cross-ear biopotential (bottom, blue) also capture the QRS complex. However, while the cross-ear signal maintains a well-defined T wave, the single-ear trace exhibits a more attenuated and less consistent post-QRS signal, likely due to the localized nature of the single-ear measurement. Since all signals are aligned to the R-peaks, EEG interference, and other noise sources are effectively averaged out, resulting in a clear ECG waveform. However, this clarity is a direct consequence of averaging across all recordings, which suppresses noise. In a real-time inference setting, as proposed with the DeepMF-mini algorithm, such averaging is not possible, making R-peak detection a far more challenging task. Still, the aligned and averaged traces confirm that in-ear biopotentials can capture ECG waveforms, demonstrating the feasibility of R-peak detection in both single-ear and cross-ear configurations for our setup. 

\subsection{Inference examples}
\begin{figure}[tbh]
\centering
\includegraphics[width=0.85\columnwidth, trim=0cm 0cm 0cm 0cm, clip]{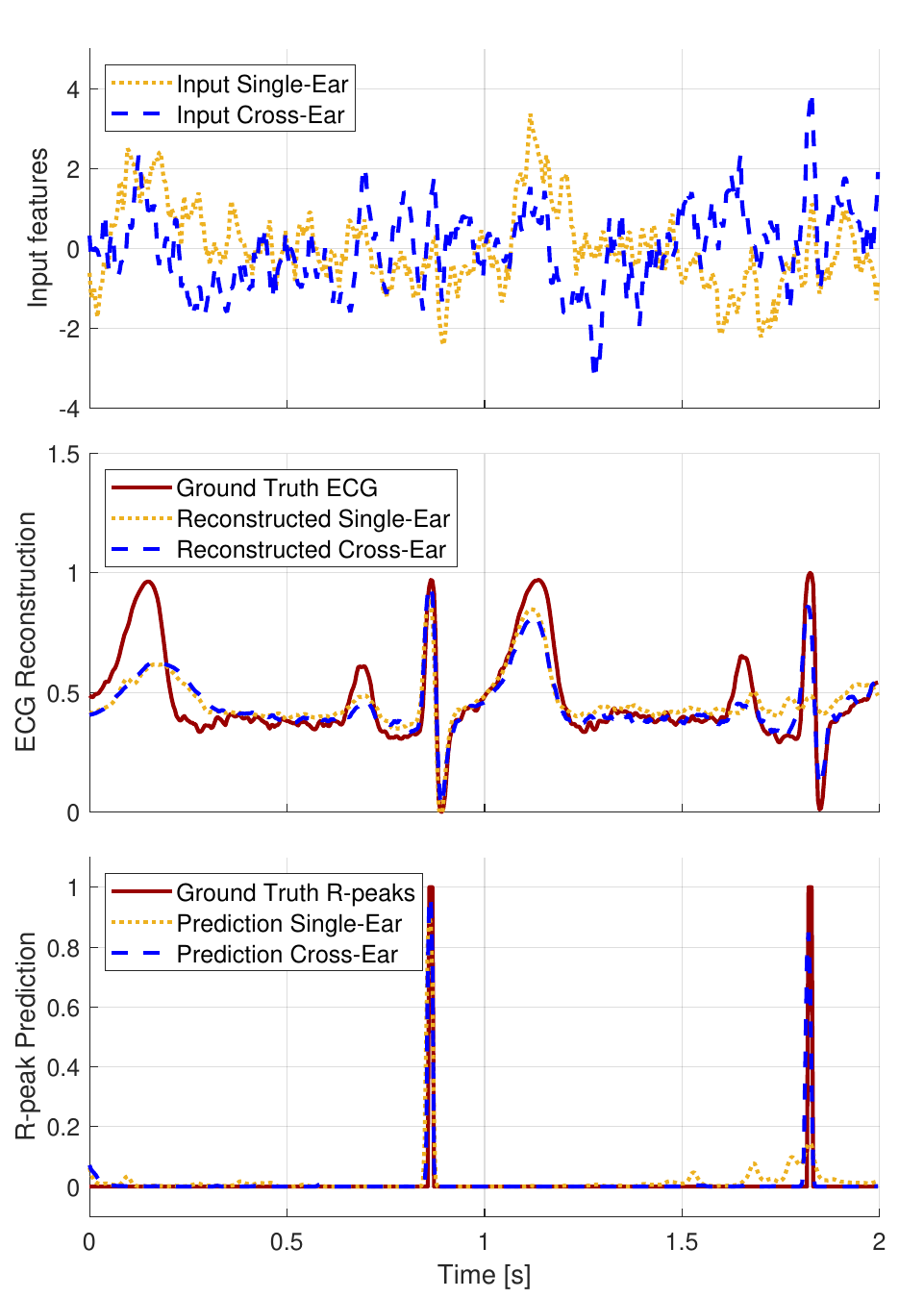}
\caption{Inference example illustrating the input signals for single-ear and cross-ear configurations (top), the corresponding ECG reconstructions alongside the ground truth ECG (middle), and the predicted R-peak locations compared to the ground truth R-peaks (bottom).}
\label{fig:sample_inference}
\vspace{-0.6cm}
\end{figure}

Fig.~\ref{fig:sample_inference} shows an example inference. Fig.~\ref{fig:sample_inference} (top) shows the filtered biopotential signals for both single-ear and cross-ear configurations, where no distinct ECG pattern is visible at this stage. Fig.~\ref{fig:sample_inference} (middle) shows the reconstructed ECG traces for both configurations alongside the ground truth ECG. The apparent large amplitude of the T-wave results from tanh scaling. It can be observed that the ECG waveform around \SI{1}{s} is accurately reconstructed by both input configurations, whereas the QRS complex around \SI{1.8}{s} is correctly reconstructed using cross-ear input but is missed by the single-ear input. This pattern of missed reconstructions, particularly near the end of the window, occurs frequently; thus, this region is excluded when inferring R-peaks. Fig.~\ref{fig:sample_inference} (bottom) displays the predicted R-peak locations alongside the ground truth R-peak positions: similar to the previous case, both cross-ear peaks are successfully detected, whereas the peaks corresponding to the single-ear input are not identified (near the window’s edge). However, due to the rolling window approach, which shifts the signal by \SI{400}{ms}, originally missed peaks at the window boundary can still be detected in subsequent iterations, improving robustness in peak detection.

\begin{figure}[tbh]
\centering
\includegraphics[width=0.9\columnwidth, trim=0cm 0.6cm 0cm 0cm, clip]{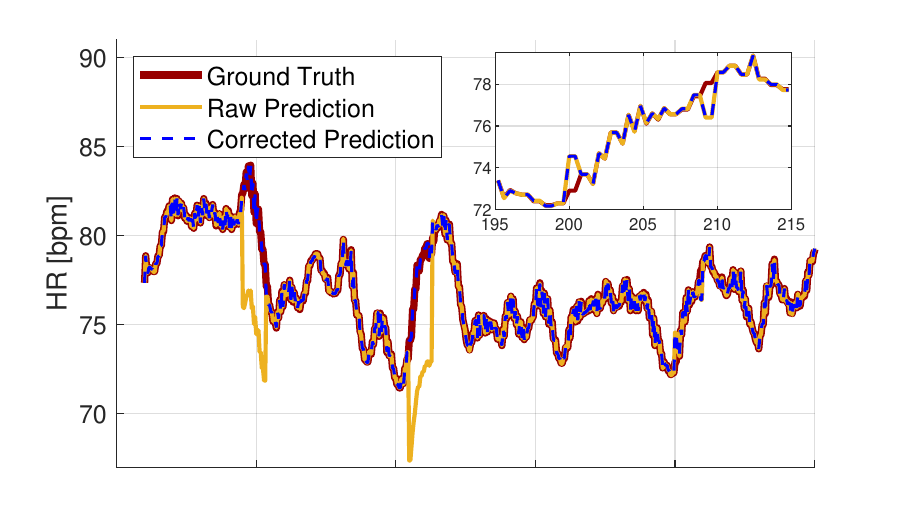}
\vspace{-0cm}
\includegraphics[width=0.85\columnwidth, trim=0cm 0cm 0cm 0cm, clip]{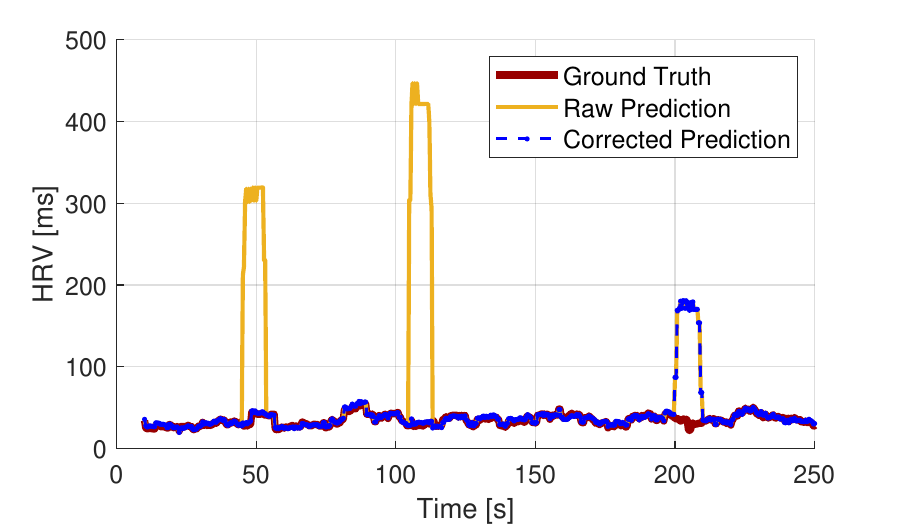}
\vspace{-0.1cm}
\caption{HR and HRV prediction.}
\label{fig:HR_HRV_curve}

\end{figure}

Fig.~\ref{fig:HR_HRV_curve} presents the rolling output of the computed HR and HRV curves over \SI{250}{s}. In Fig.~\ref{fig:HR_HRV_curve} (top), the raw and corrected predictions generally follow the ground truth. However, around \SI{50}{s} and \SI{110}{s}, the raw prediction deviates from the reference signal. The correction algorithm (Algorithm~\ref{algo:correct_HR}) successfully compensates for these missed peaks, restoring alignment with the ground truth. Between \SI{200}{s} and \SI{210}{s}, minor inaccuracies in the HR prediction are observed. These deviations, highlighted in the inset plot, remain uncorrected since they do not exhibit clear anomalies.
In Fig.~\ref{fig:HR_HRV_curve} (bottom), which displays the \ac{HRV} curves, the corrected peaks at \SI{50}{s} and \SI{110}{s} are fully restored by the correction algorithm. However, the inaccuracies in R-peak prediction between \SI{200}{s} and \SI{210}{s} propagate into HRV estimation, underscoring the sensitivity of HRV calculations, where even minor R-peak deviations can lead to significant HRV discrepancies.

\subsection{Algorithm Evaluation}
\label{subsec:Results}

\begin{figure}[tbh]
\centering \includegraphics[width=0.85\columnwidth, trim=0cm 0cm 0cm 0cm, clip]{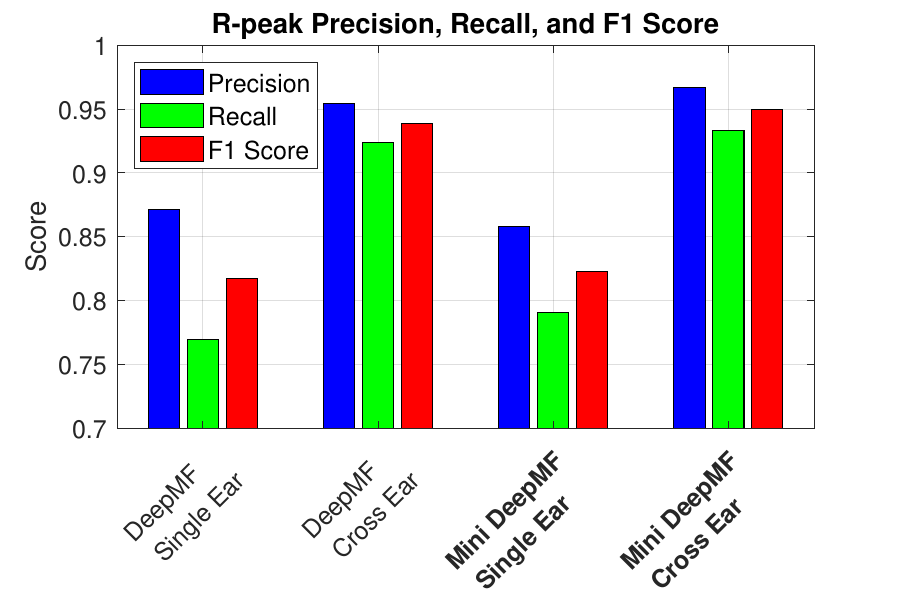}
\vspace{-0.2cm}
\caption{R-peak prediction in \ac{LOSO} \ac{CV}.} \label{fig:precision_recall_f1}
\end{figure}

Fig.~\ref{fig:precision_recall_f1} presents the achieved precision, recall, and F1-score.

The results indicate that the proposed DeepMF-mini method achieves slightly better performance in the cross-ear configuration and remains comparable to DeepMF in the single-ear configuration. The proposed DeepMF-mini algorithm reaches an overall F1-score of 0.82 and 0.95 for single-ear and cross-ear configurations, respectively.

\begin{figure}[tbh] \centering \includegraphics[width=0.85\columnwidth, trim=0cm 0cm 0cm 0cm, clip]{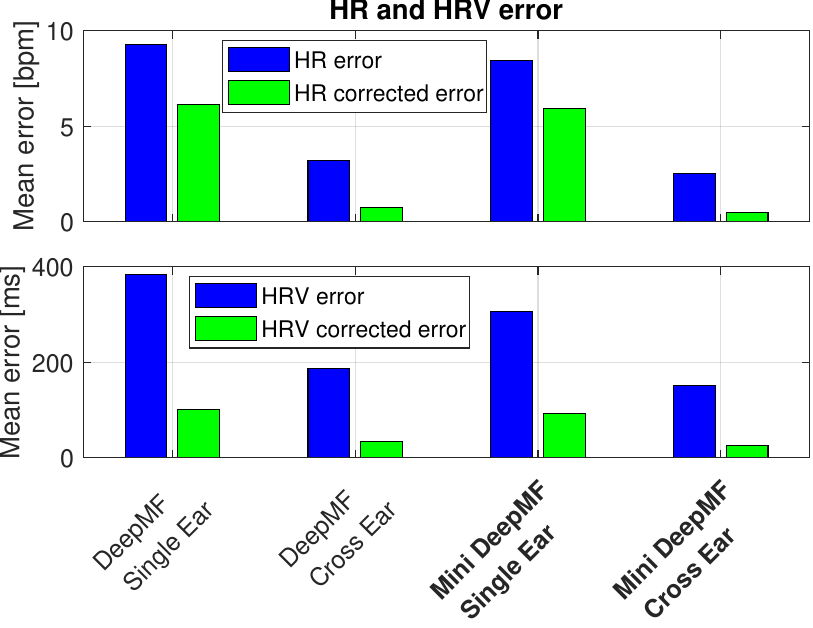}
\caption{Mean HR. and HRV error (all subjects, \ac{LOSO} \ac{CV}).} \label{fig:hr_hrv_error}
\end{figure}

Fig.~\ref{fig:hr_hrv_error} presents the error in HR and HRV predictions, both before and after correction using Algorithm \ref{algo:correct_HR}. Similar to the R-peak detection results, the DeepMF-mini algorithm demonstrates slightly better performance (with respect to Deep-MF) in the cross-ear configuration, while achieving comparable results in the single-ear configuration. Additionally, the effectiveness of Algorithm \ref{algo:correct_HR} in correcting missed or duplicate R-peaks is evident. Overall, DeepMF-mini achieves a mean HR prediction error of \SI{5.91}{bpm} and \SI{0.49}{bpm} in the single-ear and cross-ear configurations, respectively (vs. \SI{6.15}{bpm} and \SI{0.75}{bpm} DeepMF HR errors). For HRV prediction, the algorithm results in a mean error of \SI{91.57}{ms} in the single-ear setup and \SI{25.82}{ms} in the cross-ear setup (vs. \SI{100.96}{ms} and \SI{33.24}{ms} DeepMF HRV errors).

\vspace{-0.2cm}
\subsection{Edge Deployment Results}
\label{subsec:edge_deployment}

The quantized model, with a size of \SI{10}{kB}, maintained the same accuracy as its floating-point counterpart.
Table~\ref{tab:deployment} summarizes the implementation results.  At \SI{240}{MHz} and \SI{0.65}{V}, inference time is \SI{1.47}{ms}, consuming \SI{36.7}{\micro J} per inference, covering the entire GAP9 power domain. Operating at the most energy-efficient setting (NE16 @ \SI{240}{MHz}) and outputting a prediction every \SI{400}{ms} results in a total system power consumption of \SI{7.6}{mW}, including both data acquisition and edge processing.

\begin{table}[htb]
\renewcommand{\arraystretch}{0.95}
  \centering
  \caption{DeepMF-mini deployment performance}\label{tab:deployment}
  \vspace{-0.1cm}
  {
    \footnotesize
    \begin{tabular}{lp{2cm}p{2cm}}
      \toprule
      Network & \multicolumn{2}{c}{DeepMF-mini}\\
      \cmidrule(r){1-1} \cmidrule(r){2-3}
      MCU & \multicolumn{2}{c}{GAP9 (1+9$\times$RISCV) NE16}\\

      \cmidrule(r){2-3}
      MACs & \multicolumn{2}{c}{3,504,200}\\
      \cmidrule(r){2-3}
      Frequency [MHz] & 370 & 240 \\
      \cmidrule(r){1-1} \cmidrule(r){2-2} \cmidrule(r){3-3}
      Time/inference [ms]   & 0.97 & 1.47 \\
      MACs/cycle & 10.6 & 10.6 \\
      Inference power [mW] & 47.38 & 24.95 \\
      Energy/inference [uJ] & 45.8 & 36.7 \\
      En. eff. [GMAC/s/W] & 76.5 & 95.5\\
      \bottomrule
    \end{tabular}
  }
  \vspace{-0.6cm}
\end{table}

\subsection{Comparison to SoA}
\label{soa_comparison}
\vspace{-0.1cm}

Table~\ref{table:ear_ecg_comparison} compares our work to previous studies based on ear ECG measurement. Most of the prior works used cross-ear configurations: to the best of our knowledge, only \cite{YariciMetin2024Hfor} and this paper additionally investigated the use of a single-ear ECG setup. Still, while in \cite{YariciMetin2024Hfor}, authors assessed the general feasibility of ECG measurements from the ear via ECG waveform averaging, our work, instead, novels in providing a method for real-time R-peak and HR/HRV derivation.


\begin{table*}[h!]
\begin{center}
\begin{threeparttable}[b]
\caption{Comparison of Ear ECG Measurement Approaches}
\label{table:ear_ecg_comparison}
\scriptsize{
\begin{tabular}
{
p{0.2in}> 
{\centering\arraybackslash}p{0.42in}> 
{\centering\arraybackslash}p{0.9in}> 
{\centering\arraybackslash}p{0.96in}> 
{\centering\arraybackslash}p{0.7in}> 
{\centering\arraybackslash}p{0.9in}> 
{\centering\arraybackslash}p{0.15in}> 
{\centering\arraybackslash}p{0.52in}> 
{\centering\arraybackslash}p{0.35in} 
}

\toprule[0.20em]
\textbf{Paper} & \textbf{Ear ECG Setup} & \textbf{Processing} & \textbf{Approach} & \textbf{R-peak det. precision / recall} & \textbf{HR/HRV error} & \textbf{Single Ear?} & \textbf{Wearable Platform?} & \textbf{Deploy-ment} \\ 
\midrule

\textbf{\cite{DaviesHarryJ.2024TDFR}} &
Cross-ear &
Filtering, Autoencoder, Latent space-classifier &
Simultaneous arm-ECG and ear-ECG recordings &
\SI{91.2}{\%} / \SI{94.9}{\%} & - / - & No & No BrainAmp \& Sonmo HD & No \\

\midrule

\textbf{\cite{TianHaozhe2023HHRV}} &
Cross-ear &
Filtering, template matching, Hilbert transform &
Simultaneous arm-ECG, ear-ECG, and ear-PPG recordings
&
\SI{68}{\%} / \SI{67}{\%} 
&
- / - \footnotemark[1]{}
& No & n/s BioBoard & No \\

\midrule

\textbf{\cite{zylinski_hearables_2024}} &
Cross-ear &
Autoencoder, Latent space-classifier &
Simultaneous arm-ECG and ear-ECG recordings
& - / - &
29.1bpm\footnotemark[2]{}, 8.0bpm\footnotemark[3]{} / - & No & n/s & No \\

\midrule

\textbf{\cite{OcchipintiEdoardo2024IESE}} &
Cross-ear &
Filtering, Autoencoder, Scipy findpeaks &
Simultaneous arm-ECG and ear-ECG recordings
& \SI{90.0}{\%} / -  &
4.52bpm / - & No & No BrainAmp \& Sonmo HD & No \\

\midrule

\textbf{\cite{YariciMetin2024Hfor}} &
Single-ear, cross-ear
&
Filtering, Averaging, and Ensemble 
&
Simultaneous arm-ECG \mbox{and ear-ECG recordings\footnotemark[4]{}}
&
- / - &
- / - & No & No g.USBamp & No  \\

\midrule

\textbf{This work} &
Single-ear, cross-ear &
Filtering, Autoencoder, Latent space-classifier &
Simultaneous arm-ECG and ear-ECG recordings &
85.8\%\footnotemark[5], 
96.7\%\footnotemark[6] / 
79.1\%\footnotemark[5], 
93.4\%\footnotemark[6] & 
5.91bpm\footnotemark[5], 
0.49bpm\footnotemark[6]{} \mbox{/ 
91.57ms\footnotemark[5], 
25.82ms\footnotemark[6]{}} & Yes & Yes BioGAP & Yes \SI{7.6}{mW} \\

\bottomrule
\end{tabular}
}

\begin{tablenotes}
\item \footnotemark[1]{}Compute HR and HRV features from R-peak detection but don't provide quantifications. 
\footnotemark[2]{}HR estimation from multimodal signals (ear-ECG, PPG). \footnotemark[3]{}HR estimation using only ear-ECG.
\footnotemark[4]{}Assesses the feasibility of measuring ECG from the ear region using single-ear and cross-ear configurations. Rather than performing real-time R-peak detection or HR/HRV calculation, ECG waves are averaged and aligned to the R-peak over 240 or 540 cardiac cycles.
\footnotemark[5]{}Single-ear
\footnotemark[6]{}Cross-ear
\end{tablenotes}

\end{threeparttable}
\end{center}
\vspace{-8mm}
\end{table*}

For R-peak detection, our DeepMF-mini algorithm achieves performance comparable to DeepMF \cite{DaviesHarryJ.2024TDFR} (tested with a cross ear configuration), while being significantly more lightweight (10k vs. 759k parameters) and also outperforming traditional template-matching methods.
For HR estimation, we achieve the lowest HR error (\SI{0.49}{bpm}) compared to \SI{4.52}{bpm} of \cite{OcchipintiEdoardo2024IESE} and \SI{8.0}{bpm} of \cite{zylinski_hearables_2024}. Furthermore, in this work, we additionally quantify the HVR estimation performance (with an error of \SI{25.82}{ms}), which sets a benchmark for future developments.

Finally, to our knowledge, this is the first study conducted in a fully wearable setup, successfully deploying the trained network on an edge device (GAP9 embedded in BioGAP) for real-time processing in a fully wearable device.

\section{Conclusion}
\vspace{-0.1cm}

This work presents the first real-time user-generic R-peak detection and HR/HRV calculation from a single-ear wearable ECG system, leveraging DeepMF-mini, a lightweight (10k parameters), edge-efficient adaptation of DeepMF with a dedicated prediction refinement algorithm. Our system achieves SoA performance, with an R-peak detection F1-score of 0.82 (single-ear) and 0.95 (cross-ear), and mean HR errors of \SI{5.91}{bpm} (single-ear) and \SI{0.49}{bpm} (cross-ear). For HRV estimation, it achieves mean errors of \SI{91.57}{ms} (single-ear) and \SI{25.82}{ms} (cross-ear). Deployed on BioGAP, it demonstrates fast inference (\SI{1.47}{ms}), SoA energy efficiency (\SI{36.7}{\micro J/inference}), and continuous operation for 36 hours with a \SI{75}{mAh} battery. 

However, real-world applications may encounter challenges such as motion artifacts and signal interference, potentially impacting measurement accuracy during everyday activities.
Future work will therefore focus on expanding subject diversity, enhancing artifact detection, and integrating multimodal sensing to further improve reliability in real-world settings, enabling continuous, unobtrusive cardiovascular monitoring in wearable applications.

\section*{Acknowledgment}
\vspace{-0.1cm}
We thank A. Blanco Fontao (ETH Zürich) and H. Gisler (ETH Zürich) for technical support.

\bibliographystyle{IEEEtran}
\bibliography{bibliography.bib}

\end{document}